



\font\twelverm=cmr10  scaled 1200   \font\twelvei=cmmi10  scaled 1200
\font\twelvesy=cmsy10 scaled 1200   \font\twelveex=cmex10 scaled 1200
\font\twelvebf=cmbx10 scaled 1200   \font\twelvesl=cmsl10 scaled 1200
\font\twelvett=cmtt10 scaled 1200   \font\twelveit=cmti10 scaled 1200
\font\twelvesc=cmcsc10 scaled 1200
\skewchar\twelvei='177   \skewchar\twelvesy='60


\def\twelvepoint{\normalbaselineskip=12.4pt plus 0.1pt minus 0.1pt
  \abovedisplayskip 12.4pt plus 3pt minus 9pt
  \belowdisplayskip 12.4pt plus 3pt minus 9pt
  \abovedisplayshortskip 0pt plus 3pt
  \belowdisplayshortskip 7.2pt plus 3pt minus 4pt
  \smallskipamount=3.6pt plus1.2pt minus1.2pt
  \medskipamount=7.2pt plus2.4pt minus2.4pt
  \bigskipamount=14.4pt plus4.8pt minus4.8pt
  \def\rm{\fam0\twelverm}          \def\it{\fam\itfam\twelveit}%
  \def\sl{\fam\slfam\twelvesl}     \def\bf{\fam\bffam\twelvebf}%
  \def\mit{\fam 1}                 \def\cal{\fam 2}%
  \def\sc{\twelvesc}               \def\tt{\twelvett}
  \def\sf{\twelvesf}
  \textfont0=\twelverm   \scriptfont0=\tenrm   \scriptscriptfont0=\sevenrm
  \textfont1=\twelvei    \scriptfont1=\teni    \scriptscriptfont1=\seveni
  \textfont2=\twelvesy   \scriptfont2=\tensy   \scriptscriptfont2=\sevensy
  \textfont3=\twelveex   \scriptfont3=\twelveex  \scriptscriptfont3=\twelveex
  \textfont\itfam=\twelveit
  \textfont\slfam=\twelvesl
  \textfont\bffam=\twelvebf \scriptfont\bffam=\tenbf
  \scriptscriptfont\bffam=\sevenbf
  \normalbaselines\rm}



\def\beginlinemode{\endmode
  \begingroup\parskip=0pt \obeylines\def\\{\par}\def\endmode{\par\endgroup}}
\def\beginparmode{\endmode
  \begingroup \def\endmode{\par\endgroup}}
\let\endmode=\par
{\obeylines\gdef\
{}}
\def\singlespace{\baselineskip=\normalbaselineskip}
\def\oneandathirdspace{\baselineskip=\normalbaselineskip
  \multiply\baselineskip by 4 \divide\baselineskip by 3}
\def\oneandahalfspace{\baselineskip=\normalbaselineskip
  \multiply\baselineskip by 3 \divide\baselineskip by 2}
\def\doublespace{\baselineskip=\normalbaselineskip \multiply\baselineskip by 2}

\newcount\firstpageno
\firstpageno=2
\footline={\ifnum\pageno<\firstpageno{\hfil}\else{\hfil\twelverm\folio\hfil}\fi}

\def\toppageno{\global\footline={\hfil}\global\headline
  ={\ifnum\pageno<\firstpageno{\hfil}\else{\hfil\twelverm\folio\hfil}\fi}}
\let\rawfootnote=\footnote              
\def\footnote#1#2{{\rm\singlespace\parindent=0pt\parskip=0pt
  \rawfootnote{#1}{#2\hfill\vrule height 0pt depth 6pt width 0pt}}}
\def\raggedcenter{\leftskip=4em plus 12em \rightskip=\leftskip
  \parindent=0pt \parfillskip=0pt \spaceskip=.3333em \xspaceskip=.5em
  \pretolerance=9999 \tolerance=9999
  \hyphenpenalty=9999 \exhyphenpenalty=9999 }
\def\dateline{\rightline{\ifcase\month\or
  January\or February\or March\or April\or May\or June\or
  July\or August\or September\or October\or November\or December\fi
  \space\number\year}}
\def\received{\vskip 3pt plus 0.2fill
 \centerline{\sl (Received\space\ifcase\month\or
  January\or February\or March\or April\or May\or June\or
  July\or August\or September\or October\or November\or December\fi
  \qquad, \number\year)}}


\hsize=5.5truein
\hoffset=0.5truein
\vsize=8.5truein
\voffset=0.25truein
\parskip=\medskipamount
\toppageno
\twelvepoint
\oneandathirdspace
\def\\{\cr}
\overfullrule=0pt 



\def\tutp#1{
  \rightline{\rm TUTP--#1}} 

\def\title#1{                   
   \null \vskip 3pt plus 0.3fill \beginlinemode
   \doublespace \raggedcenter {\bf #1} \vskip 3pt plus 0.1 fill}

\def\author                     
  {\vskip 3pt plus 0.1fill \beginlinemode \doublespace \raggedcenter}

\def\affil                      
  {\vskip 3pt \beginlinemode \oneandathirdspace \raggedcenter \it}

\def\abstract                   
  {\vskip 3pt plus 0.1fill \subhead {Abstract:}
   \beginparmode \narrower \oneandahalfspace }

\def\endtopmatter               
  {\vskip 3pt plus 0.1fill \endpage \body}

\def\body                       
  {\beginparmode}               

\def\head#1{                    
   \goodbreak \vskip 0.4truein  
  {\immediate\write16{#1} \raggedcenter {\sc #1} \par}
   \nobreak \vskip 3pt \nobreak}

\def\subhead#1{                 
  \vskip 0.25truein             
  {\raggedcenter {\it #1} \par} \nobreak \vskip 3pt \nobreak}

\def\beneathrel#1\under#2{\mathrel{\mathop{#2}\limits_{#1}}}

\def\refto#1{${\,}^{#1}$}       

\newdimen\refskip \refskip=0pt
\def\references         
  {\head{References}    
   \beginparmode \frenchspacing \parindent=0pt \leftskip=\refskip
   \parskip=0pt \everypar{\hangindent=20pt\hangafter=1}}

\gdef\refis#1{\item{#1.\ }}                     

\gdef\journal#1, #2, #3 {               
    {\it #1}, {\bf #2}, #3.}            




\def\endreferences{\body}

\def\figurecaptions             
  {\endpage \beginparmode \head{Figure Captions}
   \parskip=3pt \everypar{\hangindent=20pt\hangafter=1} }

\def\endpage                    
  {\vfill\eject}

\def\endpaper   {\endmode\vfill\supereject}
\def\endjnl     {\endpaper\end}


\def\ref#1{ref.{#1}}                    
\def\Ref#1{Ref.{#1}}                    
\def\[#1]{[\cite{#1}]}
\def\cite#1{{#1}}


\def\(#1){(\call{#1})}
\def\call#1{{#1}}
\def\frac#1#2{{#1 \over #2}}

\def\12{{1\over2}}

\def\sla{\raise.15ex\hbox{$/$}\kern-.57em}
\def\leaderfill{\leaders\hbox to 1em{\hss.\hss}\hfill}
\def\twiddle{\lower.9ex\rlap{$\kern-.1em\scriptstyle\sim$}}
\def\bigtwiddle{\lower1.ex\rlap{$\sim$}}
\def\gtwid{\mathrel{\raise.3ex\hbox{$>$\kern-.75em\lower1ex\hbox{$\sim$}}}}
\def\ltwid{\mathrel{\raise.3ex\hbox{$<$\kern-.75em\lower1ex\hbox{$\sim$}}}}
\def\square{\kern1pt\vbox{\hrule height 1.2pt\hbox{\vrule width 1.2pt\hskip 3pt
   \vbox{\vskip 6pt}\hskip 3pt\vrule width 0.6pt}\hrule height 0.6pt}\kern1pt}
\def\tdot#1{\mathord{\mathop{#1}\limits^{\kern2pt\ldots}}}

\def\pmb#1{\setbox0=\hbox{#1}%
  \kern-.025em\copy0\kern-\wd0
  \kern  .05em\copy0\kern-\wd0
  \kern-.025em\raise.0433em\box0 }

\catcode`@=11
\newcount\r@fcount \r@fcount=0
\newcount\r@fcurr
\immediate\newwrite\reffile
\newif\ifr@ffile\r@ffilefalse
\def\w@rnwrite#1{\ifr@ffile\immediate\write\reffile{#1}\fi\message{#1}}

\def\writer@f#1>>{}
\def\referencefile{
  \r@ffiletrue\immediate\openout\reffile=\jobname.ref%
  \def\writer@f##1>>{\ifr@ffile\immediate\write\reffile%
    {\noexpand\refis{##1} = \csname r@fnum##1\endcsname = %
     \expandafter\expandafter\expandafter\strip@t\expandafter%
     \meaning\csname r@ftext\csname r@fnum##1\endcsname\endcsname}\fi}%
  \def\strip@t##1>>{}}

\def\citeall#1{\xdef#1##1{#1{\noexpand\cite{##1}}}}
\def\cite#1{\each@rg\citer@nge{#1}}     

\def\each@rg#1#2{{\let\thecsname=#1\expandafter\first@rg#2,\end,}}
\def\first@rg#1,{\thecsname{#1}\apply@rg}       
\def\apply@rg#1,{\ifx\end#1\let\next=\relax
\else,\thecsname{#1}\let\next=\apply@rg\fi\next}

\def\citer@nge#1{\citedor@nge#1-\end-}  
\def\citer@ngeat#1\end-{#1}
\def\citedor@nge#1-#2-{\ifx\end#2\r@featspace#1 
  \else\citel@@p{#1}{#2}\citer@ngeat\fi}        
\def\citel@@p#1#2{\ifnum#1>#2{\errmessage{Reference range #1-#2\space is bad.}
    \errhelp{If you cite a series of references by the notation M-N, then M and
    N must be integers, and N must be greater than or equal to M.}}\else%
 {\count0=#1\count1=#2\advance\count1
by1\relax\expandafter\r@fcite\the\count0,%

  \loop\advance\count0 by1\relax
    \ifnum\count0<\count1,\expandafter\r@fcite\the\count0,%
  \repeat}\fi}

\def\r@featspace#1#2 {\r@fcite#1#2,}    
\def\r@fcite#1,{\ifuncit@d{#1}          
    \expandafter\gdef\csname r@ftext\number\r@fcount\endcsname%
    {\message{Reference #1 to be supplied.}\writer@f#1>>#1 to be supplied.\par
     }\fi%
  \csname r@fnum#1\endcsname}

\def\ifuncit@d#1{\expandafter\ifx\csname r@fnum#1\endcsname\relax%
\global\advance\r@fcount by1%
\expandafter\xdef\csname r@fnum#1\endcsname{\number\r@fcount}}

\let\r@fis=\refis                       
\def\refis#1#2#3\par{\ifuncit@d{#1}
    \w@rnwrite{Reference #1=\number\r@fcount\space is not cited up to now.}\fi%
  \expandafter\gdef\csname r@ftext\csname r@fnum#1\endcsname\endcsname%
  {\writer@f#1>>#2#3\par}}

\def\r@ferr{\endreferences\errmessage{I was expecting to see
\noexpand\endreferences before now;  I have inserted it here.}}
\let\r@ferences=\references
\def\references{\r@ferences\def\endmode{\r@ferr\par\endgroup}}

\let\endr@ferences=\endreferences
\def\endreferences{\r@fcurr=0
  {\loop\ifnum\r@fcurr<\r@fcount
    \advance\r@fcurr by 1\relax\expandafter\r@fis\expandafter{\number\r@fcurr}%
    \csname r@ftext\number\r@fcurr\endcsname%
  \repeat}\gdef\r@ferr{}\endr@ferences}


\let\r@fend=\endpaper\gdef\endpaper{\ifr@ffile
\immediate\write16{Cross References written on []\jobname.REF.}\fi\r@fend}

\catcode`@=12

\citeall\refto          
\citeall\ref            %
\citeall\Ref            %

\vglue 0.5 truein
\tutp{-92-12}

\title
{
Electroweak Strings: A Progress Report\footnote*{Talk given at
Texas/Pascos 1992 at Berkeley.}
}
\smallskip
\author
{Tanmay Vachaspati}
\affil
{
Tufts Institute of Cosmology, Department of Physics and Astronomy,
Tufts University, Medford, MA 02155.
}

\medskip


\body

The standard model for the electroweak interactions
is known\refto{yn, nm, tv, tvmb}
to contain vortex solutions that are the
usual Nielsen-Olesen $U(1)$ vortices\refto{hnpo} embedded
in the larger electroweak symmetry group $SU(2)\times U(1)$.
After briefly describing these vortex solutions, I will discuss
their stability and argue that the presence of bound states
can stabilize them.
Then I will draw an intimate
connection between the sphaleron\refto{nm} and electroweak strings
and show how the sphaleron can be reinterpreted as a segment
or a loop of electroweak string\refto{mbmbtv}.

In the notation of Ref. \cite{jct}, the
three linearly independent electroweak string
solutions are the $W^1$ string,
$$
\Phi = f_{NO} (r) e^{i\tau^1 \theta} \pmatrix{0\cr 1\cr} ,\ \ \
W_\mu ^1 = - {{v_{NO} (r)} \over r} {\hat e}_\theta
\eqno (1)
$$
the $W^2$ string,
$$
\Phi = f_{NO} (r) e^{i\tau^2 \theta} \pmatrix{0\cr 1\cr} ,\ \ \
W_\mu ^2 = - {{v_{NO} (r)} \over r} {\hat e}_\theta
\eqno (2)
$$
and the $Z$ string,
$$
\Phi = f_{NO} (r) e^{iT \theta} \pmatrix{0\cr 1\cr} ,\ \ \
Z_\mu  = - {{v_{NO} (r)} \over r} {\hat e}_\theta \ .
\eqno (3)
$$
In the above equations, $(r, \theta ,z)$ are cylindrical coordinates,
the subscript $NO$ stands for the solution that Nielsen-Olesen
found and the fields not explicitly shown are zero. (For example,
in the $Z$ string, $W_\mu ^1 = 0 = W_\mu ^2 = A_\mu$.) The matrix
$T$ is the generator associated with the $Z$ gauge field:
$T = diag( -cos2\theta_W , 1 )$.

Of these three solutions, the $W^1$ string is gauge equivalent to
the $W^2$ string.
The reason for listing both the $W^1$ and
$W^2$ strings separately is that one may have a loop of string
which consists of segments of $W^1$ string and $W^2$ string and such
a loop is not gauge equivalent to a loop of pure $W^1$
or pure $W^2$ string.

The vortex solutions in the electroweak model are not topological
and so their stability is not guaranteed. Indeed, extensive
stability analyses have shown\refto{mjlptv, wp} that the $Z$ string is
stable only in a region very close to $sin^2 \theta_W = 1$. Simple
analytic arguments have also shown\refto{mbmbtv} that
the $Z$ string is unstable
when $sin^2 \theta_W = 0.5$ and
that $W$ strings are unstable. On the other hand,
it has been shown\refto{tvrw} that the presence of bound states on
the strings can dramatically improve their stability. It is this issue
that we discuss first.

The electroweak model contains leptons and quarks in addition to the
bosonic fields that make up the strings. In any realistic setting -
such as the early universe -
where we expect the formation
of strings, these other fermionic fields will also be present and
so it would be incorrect to consider the properties of the string
solutions in the complete absence of these fields. In extensions
of the standard electroweak model, one has additional scalar
fields too. So the question
is if the other fields can make a difference to the stability of
the string.

To answer this question we note that the leptons and quarks carry
various conserved charges and also get a mass by the Higgs mechanism
when the Higgs field acquires a vacuum expectation value. (The neutrinos
remain massless and so the following discussion does not apply to them.)
Therefore these particles are massless inside the string where the
Higgs field vanishes and are massive outside the string where the
Higgs field acquires its vacuum expectation value. To be specific,
consider putting
an electron in the vicinity of a string. The electron experiences a
potential well since it is massless inside the string and massive
outside. This potential well is in two dimensions and will always have
a bound state. Hence the electron will form a bound state with the
string\refto{zeromodes}. Then we could add more electrons to the string
and construct a string ``atom''. The total number of electrons
that can be put in this atom
will be limited by the repulsion between the electrons.
But it is not necessary to consider the case where we only put a single
species of particle on the string. For example, we could put electrically
neutral
combinations of quarks and leptons on the string and then it would
be possible to increase the number of bound states. If we are thinking
of some extension of the standard electroweak model, we also have
the possibility of dressing the string with scalar field bound states.

Let us
now examine the stability of the string in the presence of such
bound states.
{}From the perspective of the particle that is bound to the string,
it is unfavorable for the string to decay because then the particle
would have to acquire a mass. Therefore, the bound particles have a
tendency to maintain the string configuration and, hence, stabilize
the string . Recall that this
is the very argument that allows for the existence of non-topological
solitons\refto{tl} or even for the stability of ordinary atomic nuclei.

The above arguments show that by putting particles on the electroweak
strings, they can be made more stable. To improve the stability further,
one would have to keep on adding more and more of the conserved charge
that the particles carry.
But can any string solution
be stabilized in this way? This is a dynamical question and the answer
will depend on the charges and the masses of the particles,
the coupling constants and the precise particle content that is put
on the string. But to get an idea of how strong the effect of bound
states can be, it is good to consider a simple toy model in which
the electroweak
theory has an additional complex scalar field that does
not couple to the gauge fields. Such a field does not exist in the
standard electroweak model but it could arise in an extension.

The Lagrangian for the electroweak model with the extra scalar field,
$\chi$ is,
$$
L = L_{ew} + |\partial_\mu \chi |^2 - \lambda_2 |\chi |^4
                      - \lambda_3 (\Phi^{\dag} \Phi \pm m^2 ) |\chi |^2
\eqno (4)
$$
where $L_{ew}$ is the standard electroweak Lagrangian. If the minus sign
is chosen in front of $m^2$, some constraints on the coupling constants
must be satisfied for the true vacuum to be at $\chi = 0$.

The model now has an extra global $U(1)_{gl}$ symmetry in addition to the
usual $SU(2)\times U(1)$ gauged symmetry. The global charge corresponding
to this extra symmetry is conserved and $\chi$ bound states carry a fixed
amount of $U(1)_{gl}$ charge. The stability of the $Z-$string with
$\chi$ bound states was analyzed in detail in
Ref. \cite{tvrw} and the result is shown in Fig. 1 for some
generic values of the parameters. The salient features
of the plot in Fig. 1 are that when there are no bound states (zero
charge on the string) the smallest value of $sin^2 \theta_W$ for which
we find stability is roughly $0.92$. Then, as the charge
on the string is increased,
the critical value of $sin^2 \theta_W$ drops very rapidly at first
and then levels off. After this stage, adding more charge only slightly
improves the stability. The smallest value of $sin^2 \theta_W$ at which
we got stability was 0.46 for our choice of parameters. We feel that
a more thorough exploration of parameter space would have yielded even
smaller values of $sin^2 \theta_W$ but did not pursue this question
since what we are investigating is, after all, only a toy model.

There are some special features of the toy model in (4) that are absent
from the standard electroweak model (but may be present in extensions
of it). The main such feature is that since the force between two
$U(1)_{gl}$ charges is mediated only by spin 0 particles, $\chi$ charges
attract. Due to this feature there is an instability in the $\chi$
distribution along the string - the linear distribution of charge is
unstable to clumping up in spherical lumps\refto{ecrkkl} and
the cylindrical string is unstable to forming spherical bulges.
There are two ways around this instability. The first way is to
realize that this instability only involves long wavelength perturbations
along the string\refto{ecrkkl} and so it would not be present if we only
have relatively short segments or small loops of string. Indeed, the
production of long metastable strings
is expected to be exponentially suppressed
in any production mechanism and so there is a good chance that there
is no need to worry about the clumping instability.
The second way out is if
the force between the particles on the string is also
mediated by vector bosons and is repulsive
or only very weakly attractive. Then the linear distribution of charge
is preferred over the spherical distribution and the clumping instability
is absent. (On the other hand, it is easier to keep on adding charge to
the string if the force between charges is attractive and so one would
expect to be able to pack less charge on the string when the force between
charges is repulsive.)

To summarize the above discussion: it is clear that bound states
dramatically improve the stability
of electroweak strings but it is yet to be seen if the standard
model with $sin^2 \theta_W = 0.23$
or some extension of it contains stable string solutions.

I would now like to discuss another unstable solution contained in the
standard electroweak model - the sphaleron\refto{nm}. The sphaleron
solution when $sin^2 \theta_W \approx 0$ is\refto{fknm}:
$$
\Phi = f(r) \pmatrix{ sin\theta e^{i\phi}\cr cos\theta\cr }
\equiv f(r) U \pmatrix{0\cr 1\cr}, \ \ \
A_\mu \equiv W_\mu ^a \tau^a = -i v(r) (\partial_\mu U ) U^{-1}
\eqno (5)
$$
where we are now using spherical coordinates and
$\tau^a$ are the Pauli spin matrices. There will also be an electromagnetic
field accompanying this field configuration and it
can be calculated\refto{fknm}
to first order in $\theta_W$. The back-reaction of the electromagnetic
field on the sphaleron configuration is second order in $\theta_W$ and
can be ignored in the limit that we are considering.

Now the matrix
$U$ is a unitary matrix and may be written as:
$$
U = exp[ i {\hat n} \cdot {\vec \tau} ]
\eqno (6)
$$
where, ${\hat n} = (sin\phi , cos\phi , 0)$. Therefore the sphaleron
Higgs configuration is that of a $W$ string loop in the $xy-$plane
provided the $\theta$
angle in eqs. (1) and (2) is taken to run from 0 to $\pi$. The loop,
however, does not consist of a pure $W^1$ or $W^2$ string; instead
at $\phi = 0$ it is made of $W^2$ string,
at $\phi = \pi /2$ it is $W^1$ string and so on.
The gauge
field configuration is chosen to be such that the fields at infinity
are pure gauge. This leads to the first interpretation of
the sphaleron\refto{mbmbtv}:
the sphaleron is a circular loop of $W^1 - W^2$ string that has collapsed to
zero radius. A consequence of this interpretation is that if we were to
start with such a loop but of a larger radius in the $xy-$plane,
it would collapse into
the sphaleron configuration of (5).

Another interpretation of the sphaleron is possible
if we consider the $\theta = \pi /2$ section of the sphaleron. In this
section, the Higgs configuration is that of the $Z-$string (except for
a trivial global rotation) in (3). Therefore if we were to stretch
the sphaleron along the $z-$axis, we would get a segment of $Z-$string.
But we know that the $Z-$string has to end on magnetic monopoles\refto{yn}
and so the sphaleron can be viewed as a collapsed segment of $Z-$string in
which a monopole and anti-monopole sit adjacent to one another.
This interpretation
makes it immediately obvious that the sphaleron has to have a
magnetic dipole moment.

The connection of the sphaleron with the magnetic monopole pair can
be seen directly from the Higgs configuration for the monopole\refto{yn}:
$$
\Phi = {\bar f}(r) \pmatrix{ e^{i\phi} sin\theta /2\cr cos\theta /2\cr } \ .
\eqno (7)
$$
The sphaleron Higgs configuration in (5) covers the entire monopole
configuration when $\theta$ in the sphaleron goes from $0$ to $\pi /2$.
And when we take $\theta$ to go from $\pi /2$ to $\pi$ in the sphaleron,
we cover the anti-monopole configuration. Therefore the
northern hemisphere of the sphaleron configuration describes a magnetic
monopole while the southern hemisphere describes an anti-monopole.
The monopole-anti-monopole pair are connected by a $Z-$string which
runs parallel to the $z-$axis and is of zero length. This is the
second interpretation of the sphaleron\refto{mbmbtv}.
A consequence is that, if
segments of $Z-$string were produced during, say, the electroweak
phase transition, these would collapse into the sphaleron configuration.

The sphaleron is a saddle-point solution in the electroweak model and
the unstable direction has been described in Ref. \cite{nm}. The
above interpretations of the sphaleron describe the stable directions
in configuration space. That is, if the sphaleron is perturbed to
a higher energy configuration in certain directions
- in directions that ``go up the saddle'' -
it would deform into
segments of $Z-$string or loops of $W-$string.

The sphaleron plays an important role in baryon number violating processes
at high temperatures and so we expect that electroweak strings could play
a similar role. However, the energy of the sphaleron is
the smallest of all such configurations and
hence can be expected to play the dominant role in the production or
dissipation of baryon number.

As discussed above, there is a possibility that loops and
segments of strings can be stabilized by the presence of bound states
on the strings.
The same mechanism might work for the sphaleron in
which case one would have a metastable particle configuration in the
electroweak model. The eventual decay of this particle would lead to the
production of baryon number.
While we do not know for certain if the stabilizing mechanism will
work for Nature's choice of parameters, it is quite likely that the
sphaleron will require a significant amount of charge to stabilize
it. In the context of the early universe, the cross-section for forming
bound states on the sphaleron could be considerably enhanced if
segments or loops of string could slowly collapse and sweep up the
required amount of charge into the sphaleron
configuration. In that case, the problem of packing the sphaleron with
a sufficient amount of charge would seem to be alleviated.

\noindent {\it{Acknowledgements:}}

This work was supported in part by the National Science Foundation.

\references

\refis{tv} Vachaspati, T. 1992. Phys. Rev. Lett. {\bf 68}: 1977;
Vachaspati, T. 1992. Nucl. Phys. B, in press.

\refis{jct} Taylor, J. C. 1976.  ``Gauge Theories of Weak Interactions'',
Cambridge University Press.

\refis{hnpo} Nielsen, H. B. \& P. Olesen. 1973. Nucl. Phys. B{\bf{61}}: 45.

\refis{yn} Nambu, Y. 1977. Nucl. Phys. B{\bf 130}: 505;
Huang, K. \& R. Tipton. 1981. Phys. Rev. D{\bf 23}: 3050.

\refis{wp} Perkins, W. 1992. University of Sussex preprint.

\refis{nm} Manton, N. S. 1983. Phys. Rev. D{\bf 28}: 2019.

\refis{tvmb} Vachaspati, T. \& M. Barriola. 1992.
Phys. Rev. Lett. {\bf 69}: 1867.

\refis{mbmbtv} Barriola, M., M. Bucher \& T. Vachaspati, in preparation.

\refis{tvrw} Vachaspati, T. \& R. Watkins. 1992. TUTP-92-10.

\refis{mjlptv} James, M., L. Perivolaropoulos \& T. Vachaspati. 1992. Phys.
Rev. D{\bf 46}: R5232.

\refis{tl} Lee, T. D. 1976. {\it In} Proceedings of the Conference on Extended
Systems in Field Theory [Phys. Rep. {\bf 23C}: 254].

\refis{ecrkkl} Copeland, E., R. Kolb \& K. Lee. 1988. Phys. Rev. D{\bf 38}:
3023.

\refis{fknm} Klinkhammer, F. R. \& N. S. Manton. 1984. Phys. Rev. D{\bf 30}:
2212.

\refis{zeromodes} This is similar to what happens when we have zero modes
on the string: Jackiw, R. \& P. Rossi. 1981. Nucl. Phys. B{\bf 190} [FS3]:
681.

\endreferences


\medskip

\beginsection{Figure Caption}

The value of $sin^2 \theta_W$ above which strings with a certain
amount of charge per unit length
$\bar q$ are stable plotted versus $\bar q$ for some
generic values of the parameters.

\vfill
\eject

\endjnl
\end